\documentclass[prc,preprint,aps,showpacs,showkeys]{revtex4}

\usepackage{graphicx}
\usepackage{dcolumn}
\usepackage{bm}

\begin{document}
\renewcommand{\thefootnote}{\fnsymbol{footnote}}


\title{Scalar and vector form factors of the in-medium nucleon}


\author{K. Saito}
\email[]{ksaito@tohoku-pharm.ac.jp}
\affiliation{Tohoku College of Pharmacy, Sendai 981-8558, Japan}
\author{K. Tsushima}
\email[]{tsushima@physast.uga.edu}
\affiliation{Department of Physics and Astronomy, University of Georgia, 
Athens, GA 30602, USA}


\date{\today}

\begin{abstract}

Using the quark-meson coupling model, we calculate the form 
factors at  $\sigma$- and $\omega$-nucleon 
strong-interaction vertices in nuclear matter.
The Peierls-Yoccoz projection technique is used to take account of 
center of mass and recoil corrections. 
We also apply the Lorentz contraction to the internal quark wave function. 
The form factors are reduced by the nuclear medium relative to those in vacuum. 
At normal nuclear matter density and $Q^2 = 1$ GeV$^2$, the reduction 
rate in the scalar form factor is about 15\%, which is almost identical 
to that in the vector one. 
We parameterize the ratios of the form factors in symmetric nuclear 
matter to those in vacuum as a function of nuclear density and momentum transfer. 

\end{abstract}


\maketitle


%
%

The change of hadron properties in a nuclear medium is of fundamental 
importance in understanding the implication of QCD for nuclear physics. 
One of the most famous nuclear medium effects may be the nuclear EMC effect~\cite{emc}, and it has stimulated theoretical and experimental efforts to seek nuclear quark-gluon effects for almost two decades. 

Recently, the search for modification of the electromagnetic 
form factors of bound protons has been performed in polarized 
(${\vec e}, e'{\vec p}$) scattering experiments on $^{16}$O and $^4$He nuclei~\cite{eA}. The experiments measured the ratio of 
transverse to longitudinal polarization of the ejected proton, which is 
proportional to the ratio of electric to magnetic form factors of a proton.
However, conventional calculations including free-proton form factors, 
appropriate optical potentials and bound-state wave functions as well as  relativistic corrections, meson-exchange currents (MEC), isobar contributions 
and final-state interactions, fail to reproduce the observed results in 
$^4$He~\cite{eA,conv}. 
Indeed, full agreement with the experimental data was only 
obtained when, in addition to the standard nuclear calculation, a  
change in the form factors which is caused by the structure modification of 
bound proton~\cite{eA,emff}, was taken into account.  

Recent inclusive neutrino experiments on $^{12}$C at Los Alamos~\cite{nC} 
also suggest that the measured total cross section is about 
a half of the standard, relativistic shell model calculation including 
final-state interactions within the distorted 
wave impulse approximation~\cite{neut}.  
In the neutrino reaction, the charged-current vector form factors of bound nucleons are slightly enhanced, 
while the axial form factors are quenched by the nuclear medium~\cite{axial}.  
Finally, the effect of the bound nucleon form factors reduces 
the total cross section by about 8\% relative to that calculated 
with the free form factors~\cite{qmcneut}. We stress that this correction 
is caused by the change of the internal quark wave function 
at the mean-field level and hence there is no obvious double 
counting with MEC etc. 
This is a new effect which should be taken into 
account additionally to the standard nuclear corrections. 

Furthermore, the measurements of polarization transfer observables in 
exclusive (${\vec p}, 2{\vec p}$) proton knockout reactions from various 
nuclei~\cite{rcnp,stellen} again indicate that it is difficult to 
account for the measured polarization transfers within the conventional, 
relativistic distorted wave impulse approximation~\cite{mano}. 
To reproduce the measured spin 
observables, it is necessary to simultaneously reduce the scalar ($\sigma$) 
and vector ($\omega$) coupling constants and the meson masses by 
about $10 \sim 20$\%~\cite{mano}. 
In particular, the analyzing power ($A_y$), polarization ($P$) and spin 
transfer coefficient ($D_{ss'}$) are very sensitive to the change of 
$\sigma$- and $\omega$-nucleon coupling constants and their masses in 
a nuclear medium. 
These may again imply the change in the internal structure of 
bound nucleons. 

If the quark substructure of the nucleon is modified depending on the 
nuclear environment, it would leave traces in a variety of processes and observables, including various form factors. 
These modifications of bound nucleons can be successfully described within the context of the quark-meson coupling 
(QMC) model~\cite{qmc}. In the model, the medium effects arise through 
the self-consistent coupling of $\sigma$ and $\omega$ 
mesons to confined quarks, rather than to the nucleons. 
As a result, the internal structure of the bound nucleon is modified by the 
surrounding nuclear medium. 

The electromagnetic form factors of bound 
nucleons~\cite{emff} have been studied using an improved cloudy bag model (ICBM)~\cite{py,cbm}, together with the QMC model.  In the ICBM, a 
simplified Peierls-Thouless projection technique 
(the weight function $w({\vec p})$ appearing in the nucleon wave function 
is assumed to be unity) is used to account for 
center of mass (c.m.) and recoil corrections. In addition to it, a Lorentz 
contraction of the internal quark wave function is included.  
The axial form factor in nuclear matter has also been calculated in a 
similar manner~\cite{axial}.  
Furthermore, the form factors at $\sigma$- and $\omega$-nucleon 
strong-interaction vertices in a nuclear medium should also be investigated.
The change of these form factors is very significant 
in understanding how the strong interaction is modified 
in nuclear matter.  
It is also expected to play an important role in analyzing 
the polarization transfer observables 
in the exclusive (${\vec p}, 2{\vec p}$) reactions~\cite{rcnp,stellen}. 

In this Letter, we study the scalar and vector form factors at 
$\sigma$- and $\omega$-nucleon strong-interaction 
vertices in symmetric nuclear matter. 
We shall calculate these form factors using a relativistic 
constituent quark model with a harmonic oscillator 
(HO)~\cite{toki}  or 
a linearly rising (LR) confining potential~\cite{krein} and 
the Peierls-Yoccoz (PY) projection technique.  If we use 
the "minimax" principle (or the saddle point variational principle)~\cite{minimax,krein}, 
it is easy to obtain an approximate solution to the 
Dirac equation with {\em any} potential. Since we 
choose a Gaussian wave function for a confined quark as ansatz, 
it is possible to calculate 
the form factors analytically 
and thus transparent to see how the PY projection and 
the Lorentz contraction of the quark 
wave function work in the form factors. 
Instead, in this exploratory study, we do not include  
the pion cloud effect which can explicitly be treated in the ICBM.  (We will 
study this effect in a forthcoming paper.)  

In the QMC model, 
the mean-field approximation is applied to the $\sigma$ and $\omega$ meson fields, which couple to confined  
($u$ or $d$) quarks in nuclear matter. Each quark then satisfies the Dirac equation 
\begin{equation}
[ -i{\vec \alpha}\cdot{\vec \nabla} + \gamma^0 m_q^* + U_{conf}(r) ] 
\psi({\vec r}) = E_q \psi({\vec r}) , \label{dirac}
\end{equation}
where $m_q^* = m_q - g_\sigma^q {\bar \sigma}$ and $E_q = \epsilon_q 
- g_\omega^q {\bar \omega}$ with $\epsilon_q$ the quark energy. 
We take the free quark mass $m_q$ to be 
300 MeV. The mean-field values of $\sigma$ and $\omega$ mesons are 
respectively denoted by ${\bar \sigma}$ and ${\bar \omega}$, and $g_\sigma^q$ 
and $g_\omega^q$ are the corresponding quark and meson coupling constants. 
We use a confining potential of HO type, 
$U_{conf}(r) = (c/2)(1+\beta\gamma^0)r^2$, or a LR one, 
$U_{conf}(r) = (\lambda/2)(1+\beta\gamma^0)r$, where $\beta~(0\le \beta \le 1)$ controls the strength of the Lorentz vector-type potential.  
The potential strength is taken to be $c=0.04$ GeV$^3$ or 
$\lambda = 0.2$ GeV$^2$~\cite{pot}. 

Although for the LR potential the Dirac equation cannot be solved analytically, the minimax principle allows us to obtain an approximate solution 
very easily and accurately~\cite{minimax}. 
Since the Dirac Hamiltonian does not have a lower bound for 
the energy spectrum, the 
usual variational method cannot be applied. 
The minimax principle amounts to minimizing (maximizing) 
the energy expectation value of the upper (lower) component of the quark wave function with respect to variational parameters. A trial wave function for the lowest-energy state is usually chosen as 
\begin{equation}
\psi({\vec r}) = N_0 { u(r) \choose i \xi b {\vec \sigma}\cdot{\vec r} u(r) } 
\chi_s , 
\label{trial}
\end{equation}
with $N_0$ a normalization constant, $u(r) = e^{-b^2r^2/2}$ and 
$b$ and $\xi$ the variational parameters. 
These parameters are determined so as to minimize the quark energy 
$\epsilon_q$ with respect to $b$ and maximize it with respect to $\xi$. 
Note that for the HO potential with $\beta = 1$ this gives the exact solution~\cite{hopot}. 

First, we fix the parameters of the model in vacuum.  The nucleon mass 
in vacuum (${\bar \sigma}={\bar \omega}=0$) is given by 
$M_N = 3 \epsilon_q - \epsilon_0$, where 
$\epsilon_0$ accounts for corrections of c.m. and 
gluon fluctuations.  The parameter $\epsilon_0$ is fitted so as to 
obtain the free nucleon mass $M_N (= 939$ MeV).  
The minimax principle then determines the parameters $b$ and $\xi$. 
These values are given in Table~\ref{t:cc}. 

In matter, the scalar field couples to the confined 
quark and hence the quark mass changes depending on the nuclear environment. 
The nucleon mass in matter $M_N^*$ is then reduced because the $\sigma$ 
exchange induces an attractive force between nucleons. 
In an iso-symmetric nuclear matter, the total energy 
(per nucleon) at nuclear density $\rho_B$ is given by the usual 
expression in the QMC model~\cite{qmc}
\begin{equation}
E_{tot} = \frac{4}{\rho_B (2\pi)^3} \int^{k_F}
d\vec{k} \ \sqrt{M_N^{* 2} + \vec{k}^2} + \frac{m_{\sigma}^{2}}
{2\rho_B}{\bar \sigma}^2 + \frac{g_{\omega}^2}
{2m_{\omega}^{2}}\rho_B , 
\label{tote}
\end{equation}
where $m_\sigma (=550$ MeV) and $m_\omega (=783$ MeV) are respectively the $\sigma$ and $\omega$ meson masses, and $g_\omega (= 3 g_\omega^q)$ is the $\omega$-nucleon coupling constant. The values of the scalar and vector mean fields are, 
respectively, determined by self-consistency conditions: 
$(\partial E_{tot}/\partial {\bar \sigma}) =0$ and $(\partial E_{tot}/\partial {\bar \omega}) =0$. The latter condition ensures 
baryon number conservation,  while the former 
gives a transcendental equation for the scalar field in matter. 

The coupling constants are fitted so as to reproduce the nuclear 
saturation property ($E_{tot} -M_N = -15.7$ MeV) at normal nuclear 
matter density $\rho_0 (= 0.17$ fm$^{-3}$). Note that for each 
value of $\rho_B$ one has to use the minimax principle 
to obtain the in-medium 
parameters $b$ and $\xi$. The coupling constants and 
nuclear properties at $\rho_0$ are listed in 
Table~\ref{t:cc}. The $\sigma$-nucleon coupling constant $g_\sigma$ 
is defined in terms of the quark scalar density $S_N$:  
$g_\sigma = 3 g_\sigma^q S_N({\bar \sigma}=0)$, where  
$S_N({\bar \sigma}) = \int d{\vec r}\ {\bar \psi}(r)\psi(r)$. 

The wave function for a 
nucleon moving with momentum ${\vec p}$ can be 
constructed by the PY projection technique~\cite{py,weise}: 
\begin{equation}
\Psi({\vec r}_1, {\vec r}_2, {\vec r}_3; {\vec p}) = 
N({\vec p}) \int d{\vec x}\ e^{i{\vec p}\cdot{\vec x}} 
\Phi({\vec r}_1, {\vec r}_2, {\vec r}_3; {\vec x}) , 
\label{pywf}
\end{equation}
where $N({\vec p})$ is a momentum-dependent normalization constant 
\begin{equation}
[N({\vec p})]^{-2} = \int d{\vec r} \ e^{-i{\vec r}\cdot{\vec p}}
[\rho({\vec r})]^3 , 
\label{norm2}
\end{equation}
with  
\begin{equation}
\rho({\vec r}) = \int \frac{d{\vec k}}{(2\pi)^3} \ 
e^{i{\vec k}\cdot{\vec r}} |\phi({\vec k})|^2 . 
\label{rrr}
\end{equation}
Here $\phi$ is the quark wave function in momentum space. The localized 
state $\Phi$ is simply given by a product of the three individual quark 
wave function 
\begin{equation}
\Phi({\vec r}_1, {\vec r}_2, {\vec r}_3; {\vec x}) = 
\psi({\vec r}_1-{\vec x})\psi({\vec r}_2-{\vec x})\psi({\vec r}_3-{\vec x}) , 
\label{localwf}
\end{equation}
where ${\vec x}$ refers to the location of the center of the nucleon and 
${\vec r}_j~(j = 1, 2, 3)$ specifies the position of the $j$-th quark. 

Because the nucleon consists of three point-like quarks, the expectation value of an operator with respect to the nucleon wave function Eq.(\ref{pywf}) may be given by a sum of the individual quark expectation values~\cite{weise}. 
In the Breit frame, where the initial (final) momentum of the nucleon 
is taken to be $-{\vec q}/2~({\vec q}/2$) with ${\vec q}$ the momentum transfer, 
the scalar and vector form factors are respectively given by  
\begin{equation}
\Gamma_{{s\choose v}}(Q^2) = 3 [N(Q^2)]^2 
\int d{\vec r} \ e^{i{\vec q}\cdot{\vec r}} {\bar \psi}({\vec r}) 
{1\choose \gamma^0} {\tilde \psi}({\vec r}, {\vec q}) , 
\label{svffdef}
\end{equation}
where $Q^2 \equiv -q_0^2 + {\vec q}\,^2 = {\vec q}\,^2$, and we ignore a small tensor term at the $\omega$-nucleon coupling. 
${\tilde \psi}$ in Eq.(\ref{svffdef}) is represented by 
\begin{equation}
{\tilde \psi}({\vec r}, {\vec q}) = \int \frac{d{\vec k}}{(2\pi)^3} \ 
e^{i{\vec k}\cdot{\vec r}} \phi({\vec k})W({\vec k}, {\vec q})  , 
\label{wf1}
\end{equation}
where 
\begin{equation}
W({\vec k}, {\vec q}) = 
\int d{\vec r}\ e^{-i({\vec q}/2+{\vec k})\cdot{\vec r}}
[{\bar \rho}({\vec r})]^2 , 
\label{wf2}
\end{equation}
and 
\begin{equation}
{\bar \rho}({\vec r}) = \int \frac{d{\vec k}}{(2\pi)^3} \ 
e^{-i{\vec k}\cdot{\vec r}} {\bar \phi}({\vec k})\phi({\vec k}) . 
\label{rrs}
\end{equation}

Now we can calculate the scalar and vector form factors 
in nuclear matter 
analytically: 
\begin{equation}
\Gamma_{{s\choose v}}(Q^2, \rho_B) = 
\left( \frac{Z_0(\xi^2)}{Y_0^v(\xi^2)} \right) e^{-x^2/6} 
\frac{\sum_{i=0}^2 (x^2)^i Y_i^{{s\choose v}}(\xi^2)}{
\sum_{i=0}^3 (x^2)^i Z_i(\xi^2)} , 
\label{gamma}
\end{equation}
where $x^2 = Q^2 / b^2$, 
\begin{eqnarray}
Z_0(\xi^2) &=& 1 + 3 \xi^2 + \frac{7}{2} \xi^4 + \frac{25}{18} \xi^6 , 
\label{z0} \\
Z_1(\xi^2) &=& \frac{1}{12} \xi^2 + \frac{1}{9} \xi^4 + \frac{13}{216} \xi^6 , 
\label{z1} \\
Z_2(\xi^2) &=& \frac{1}{432} \xi^4 + \frac{1}{1296} \xi^6 , 
\label{z2} \\
Z_3(\xi^2) &=& \frac{1}{46656} \xi^6 , 
\label{z3} 
\end{eqnarray}
and 
\begin{eqnarray}
Y_0^{{s\choose v}}(\xi^2) &=& 1 - {3 \choose 1} \xi^2 + 
{\frac{7}{2} \choose \frac{-7}{6}} \xi^4 \mp \frac{25}{18} \xi^6 , 
\label{y0} \\
Y_1^{{s\choose v}}(\xi^2) &=& \frac{{9 \choose -7}}{32} \xi^2 + 
\frac{{-69 \choose 67}}{128} \xi^4 \pm \frac{335}{1152} \xi^6 , 
\label{y1} \\
Y_2^{{s\choose v}}(\xi^2) &=& \pm \frac{1}{128} \xi^4 . 
\label{y2} 
\end{eqnarray}
Recall that the variational parameters $\xi$ and $b$ (thus $x^2$), 
which appear in the quark wave function, 
depend on $\rho_B$.  We have renormalized 
the vector form factor so that $\Gamma_v = 1$ is maintained 
at zero momentum transfer.  The scalar form factor is 
also rescaled by the same factor as in the vector case~\cite{weise}. 

In contrast, if the c.m. correction is ignored, 
the form factors are given by 
\begin{equation}
\Gamma_{{s\choose v}}^0(Q^2, \rho_B) = 
\frac{e^{-x^2/4}}{\left( 1 + \frac{3}{2} \xi^2 \right)} 
\left[ 1 \mp \frac{3}{2} \xi^2 \left( 1 - \frac{1}{6} x^2 \right) \right] . 
\label{gamma0}
\end{equation}
Because $x^2$ is small and $\xi \alt 0.5$ (for $\rho_B/\rho_0 \le 2.0$) 
at small momentum transfer, we can expand the form factors.  Up to 
${\cal O}(x^2)$ or ${\cal O}(\xi^2)$,  
we find that $\Gamma_{{s\choose v}}^0 = 1 - {3 \choose 0}\xi^2 -x^2/4$, while 
Eq.(\ref{gamma}) gives 
$\Gamma_{{s\choose v}} = 1 - {2 \choose 0}\xi^2 -x^2/6$. 
The c.m. correction thus moderates the reduction of the form factors.  

Apart from the c.m. correction, it is also vital to include 
the Lorentz contraction of the internal quark wave function 
at moderate or large momentum transfer~\cite{lor,py}. 
The full form factors ${\tilde{\Gamma}_{{s\choose v}}}$ can be obtained 
through a simple rescaling~\cite{py,emff}: 
\begin{equation}
\Gamma_{{s\choose v}}(Q^2) 
\to {\tilde \Gamma}_{{s\choose v}}(Q^2) = 
\eta^* \Gamma_{{s\choose v}}(\eta^*Q^2) , 
\label{svff}
\end{equation}
where $\eta^* = (M_N^*/E_N^*)^2$ with $E_N^* = \sqrt{M_N^{*2}+Q^2/4}$. 
The scaling factor in the argument arises from the coordinate transformation 
of the struck quark and the prefactor $\eta^*$ comes from the reduction of the 
integral measure of two spectator quarks in the Breit frame~\cite{py,emff}. 
Thus, the scaling factor $\eta^*$ (in vacuum $\eta$ with $M_N$) should 
appear in any nucleon(baryon)-meson form factors 
if the nucleon (baryon) is assumed to have a three-quark cluster structure. 

\begin{figure*}
\includegraphics[height=11cm]{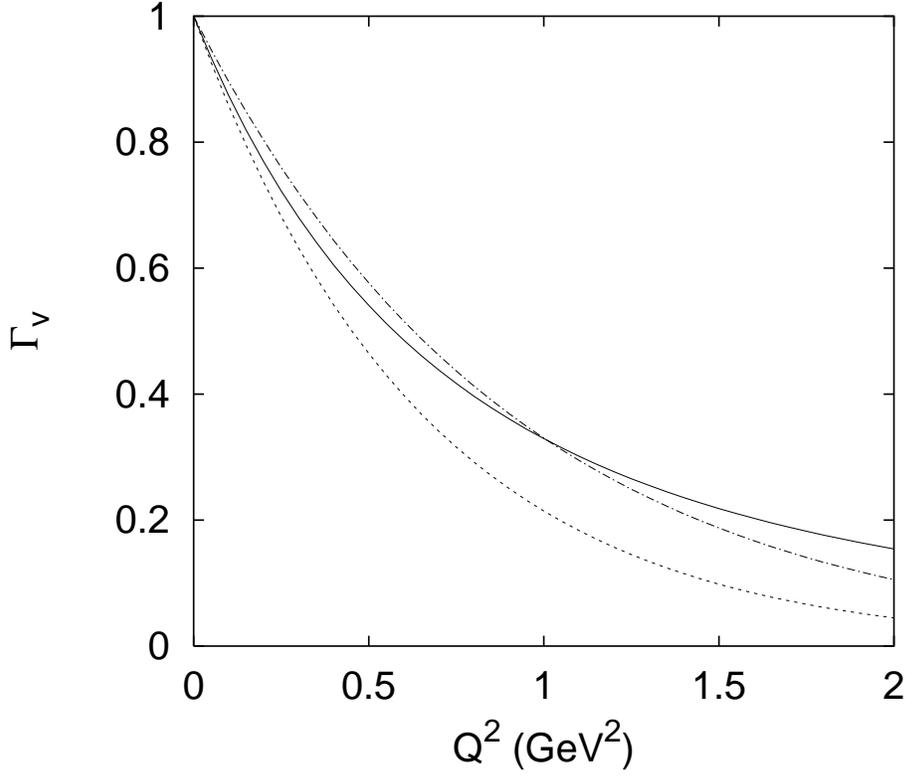}%
\caption{\label{f:v0} Vector form factor in vacuum 
(for the HO case with $\beta = 0.5$).  
The full result is denoted by the solid curve, while the dot-dashed 
curve shows the result with the c.m. correction but without 
the Lorentz contraction. The result without both corrections 
(Eq.(\protect\ref{gamma0})) is denoted by the dotted curve. }
\end{figure*}

To illustrate the effects of the c.m. correction and Lorentz contraction 
on the form factors, we show in Fig.~\ref{f:v0} the vector form factor 
in vacuum. 
The c.m. correction considerably enhances the form 
factor in comparison with the result without both effects 
(see the dotted and dot-dashed  curves in the figure). 
The effect of Lorentz contraction is also important. 
If the Lorentz contraction is ignored, the form factor drops away like 
$\sim e^{-x^2/6}$ at large $Q^2$. 
The inclusion of the Lorentz contraction removes this 
objectionable exponential falloff. 
Because of the factor $\eta$, the form factor 
is proportional to $1/(1 + Q^2/\Lambda^2)$ 
and $x^2$ is modified to $x^2/(1 + Q^2/\Lambda^2)$ 
with $\Lambda = 2M_N$ (see also Eq.(\ref{svff})). 
As a result, the inclusion of the Lorentz contraction enhances 
the form factor at large $Q^2$ (see the dot-dashed and solid curves).  

\begin{figure*}
\includegraphics[height=5.5cm]{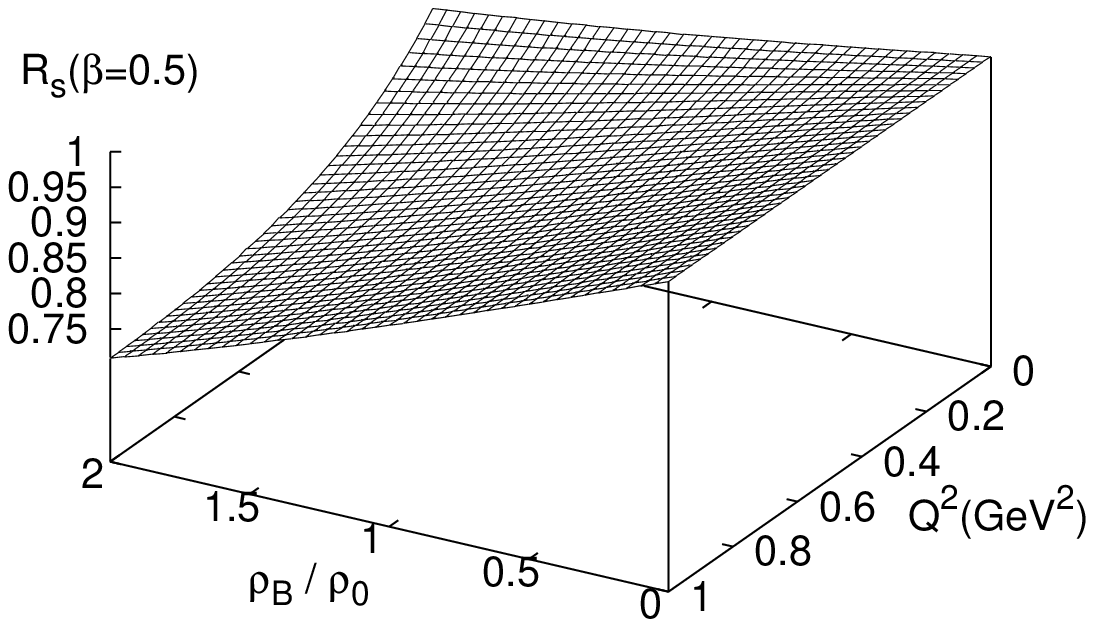}%
\includegraphics[height=5.5cm]{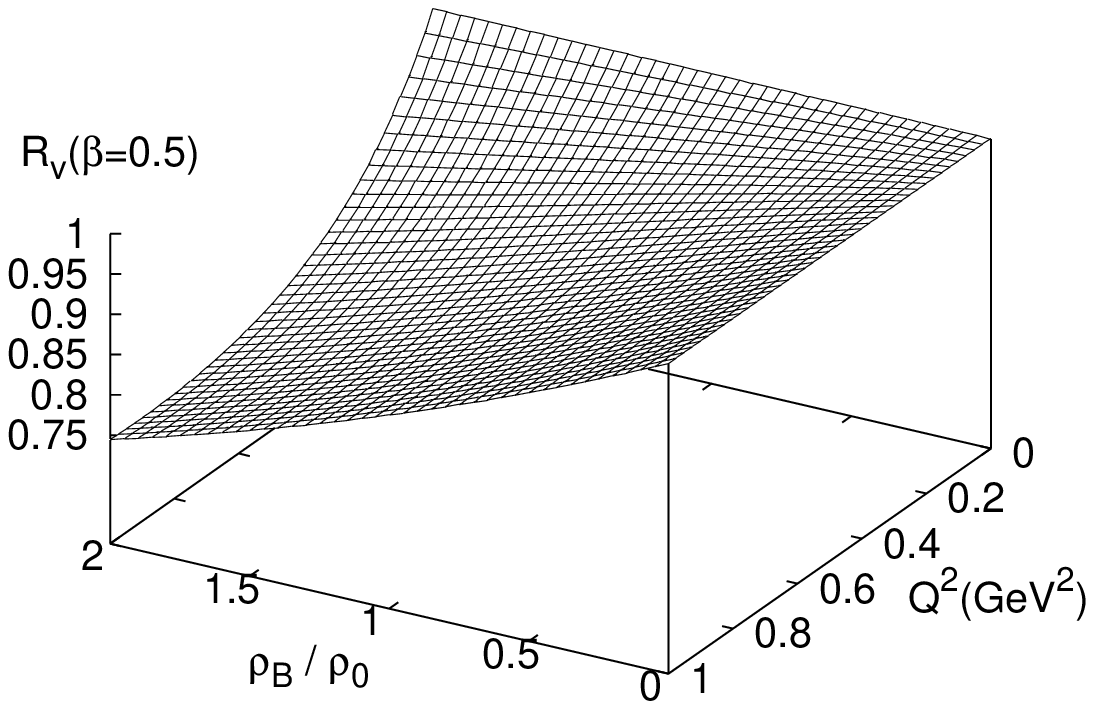}%
\caption{\label{f:hosv} Ratios for the scalar (left panel) and vector (right panel) form factors in the case of the HO potential with $\beta = 0.5$. }
\end{figure*}

Because our aim is to study the density dependence of the form factors 
in nuclear matter, we consider the ratios of the in-medium form factors to 
those in vacuum: 
\begin{equation}
R_{{s\choose v}}(Q^2, \rho_B) = 
\frac{{\tilde \Gamma}_{{s\choose v}}(Q^2, \rho_B)}{
{\tilde \Gamma}_{{s\choose v}}(Q^2, \rho_B=0)} .  
\label{ratio}
\end{equation}
The form factors in symmetric nuclear matter 
$F_{{s\choose v}}$ are thus given by 
\begin{equation}
F_{{s\choose v}}(Q^2, \rho_B) = R_{{s\choose v}}(Q^2, \rho_B) 
\times F_{{s\choose v}}^{emp}(Q^2) , 
\label{full}
\end{equation}
where $F_{{s\choose v}}^{emp}$ 
are the form factors empirically determined in 
vacuum~\cite{bonn}.
In Fig.~\ref{f:hosv}, the ratio of the in-medium scalar (vector) form 
factor to that in vacuum is illustrated as a function of $Q^2$ and 
$\rho_B$. (Because the ratios for the LR potential are similar to 
those for the HO potential, we focus on the HO case for
a while.)  At $\rho_B/\rho_0 = 1$ and $Q^2 = 1.0$ GeV$^2$, 
the in-medium scalar (vector) form factor is reduced by 15~(14)\% 
relative to that in vacuum. The reduction rate depends on $\beta$ very 
weakly. By contrast, at $\rho_B/\rho_0 = 2$ and $Q^2 = 1.0$ GeV$^2$, 
the scalar form factor decreases by 
$35~(29)~[24]$\% for $\beta = 0~(0.5)~[1.0]$, while the vector form factor 
diminishes by $28~(26)~[22]$\% for $\beta = 0~(0.5)~[1.0]$. 
At high density the dependence of the reduction on $\beta$ is thus rather 
strong, and the reduction rate is correlated with 
$M_N^*$ (see Table~\ref{t:cc}). 

As in the case of vacuum (see Fig.~\ref{f:v0}), 
the effect of Lorentz contraction is again seen at large $Q^2$. 
For example, at $\rho_B/\rho_0 = 2$ and $Q^2 = 1$ GeV$^2$, 
the vector form factor with the Lorentz contraction is about 7\% larger 
than that without it. We also note that, in the HO case with $\beta = 0.5$, 
the full vector form factor gives the root-mean-square radius of 0.53 fm. 
If we neglect the Lorentz contraction effect, it is 0.46 fm. 

Finally, we parameterize the ratios for the scalar and 
vector form factors in Eq.(\ref{full}). 
Such parameterizations are very useful in analyzing 
the experimental results, 
e.g., for the exclusive (${\vec p}, 2{\vec p}$) proton knockout 
reactions~\cite{rcnp,stellen}. With an error less than 0.2\%, 
the ratios can be represented by 
\begin{equation}
R_{{s\choose v}}(Q^2, \rho_B) = 
1 +  A_{{s\choose v}}(Q^2) (\rho_B/\rho_0) 
+ B_{{s\choose v}}(Q^2) (\rho_B/\rho_0)^2 , 
\label{approx}
\end{equation}
where  
\begin{eqnarray}
A_s(y) &=& -{0.06829 \choose 0.06323} - {0.2302 \choose 0.2464}y 
+ {0.1845 \choose 0.1711} y^2 - {0.04613 \choose 0.04072} y^3 , 
\label{As} \\
A_v(y) &=& - {0.3856 \choose 0.3738}y 
+ {0.3021 \choose 0.2668} y^2 - {0.07763 \choose 0.06494} y^3 , 
\label{Av} 
\end{eqnarray}
and 
\begin{eqnarray}
B_s(y) &=& {0.005071 \choose 0.003569} - {0.02499 \choose 0.03304}y 
+ {0.09473 \choose 0.1167} y^2 - {0.1070 \choose 0.1245} y^3 
+ {0.03914 \choose 0.04474} y^4 , 
\label{Bs} \\
B_v(y) &=& - {0.04296 \choose 0.05500}y 
+ {0.2081 \choose 0.2271} y^2 - {0.2380 \choose 0.2509} y^3  
+ {0.08967 \choose 0.09337} y^4 .  
\label{Bv} 
\end{eqnarray}
In Eqs.(\ref{As})$\sim$(\ref{Bv}), the upper (lower) numbers are for 
the case of the HO (LR) potential with $\beta = 0.5~(1.0)$, which
provides the effective nucleon mass $M_N^*/M_N = 0.71 \sim 0.72$ at $\rho_0$ (see Table~\ref{t:cc}). 
The in-medium form factors are thus given by Eqs.(\ref{full}) 
and (\ref{approx}). 

In summary, using the QMC model we have calculated the form factors 
at $\sigma$- and $\omega$-nucleon strong-interaction 
vertices in symmetric nuclear matter.  
We have applied both the PY projection technique and  
the Lorentz contraction of the internal quark wave function. 
The form factors are reduced by the nuclear medium relative to 
those in vacuum. 
The c.m. correction moderates the reduction of the form factors 
in matter, and the Lorentz contraction is vital at large momentum 
transfer. We have found that the reduction in the scalar form factor 
is about 15\% at $\rho_B/\rho_0 = 1$ and $Q^2 = 1$ 
GeV$^2$.  This rate is almost identical to that for the vector form factor. 
In contrast, the scalar and vector form factors are respectively 
reduced by about 30\% and 25\% at $\rho_B/\rho_0 =2$ and $Q^2 = 1$ GeV$^2$. 
The reduction of the form factors is expected better 
to reproduce the polarization transfer observables measured 
at RCNP and iThemba laboratory~\cite{rcnp,stellen}. 
We have parameterized the ratios of the form factors 
in symmetric nuclear matter to those in vacuum. 
This provides a convenient formula to estimate the in-medium form factors. 
It is very intriguing to re-analyze the data of polarization transfer 
observables for exclusive (${\vec p}, 2{\vec p}$) proton knockout 
reactions~\cite{rcnp,stellen} including the modification of 
both the form factors and meson masses in matter~\cite{qmc2}. 

\vspace{1.0cm}
\noindent
{\bf Acknowledgment}\\
K.T. is supported by the Forschungszentrum-J\"{u}lich,
contract No. 41445282 (COSY-058).

%

%


\newpage
\begin{table}
\caption{Coupling constants, $\epsilon_0$, $b$, $\xi$, $M_N$ and
  nuclear incompressibility $K$. 
The parameters $\epsilon_0$, $b$ and $\xi$ are fixed in vacuum, while 
$b^*$, $\xi^*$ and $M_N^*$ are calculated at normal nuclear matter density. 
Here $\epsilon_0$, $b$ and $K$ 
are quoted in GeV. The value of $\beta$ is specified in 
the parenthesis in the first column. }
\begin{ruledtabular}
\begin{tabular}{cccccccccc}
 & $g_\sigma^2$ & $g_\omega^2$ & $\epsilon_0$ & $b$ & $\xi$ & $b^*/b$ & $\xi^*/\xi$ & $M_N^*/M_N$ & $K$ \\
\hline
HO(0)   & 88.64 & 120.8 & 1.08 & 0.380 & 0.288 & 0.941 & 1.12 & 0.649 & 0.392 \\
HO(0.5) & 75.38 & 91.88 & 1.38 & 0.425 & 0.351 & 0.946 & 1.14 & 0.720 & 0.344 \\
HO(1)   & 65.12 & 69.65 & 1.63 & 0.464 & 0.401 & 0.955 & 1.15 & 0.774 & 0.316 \\
LR(0)   & 93.95 & 133.0 & 1.30 & 0.364 & 0.249 & 0.932 & 1.11 & 0.619 & 0.427 \\
LR(0.5) & 85.21 & 113.5 & 1.75 & 0.418 & 0.304 & 0.934 & 1.13 & 0.667 & 0.381 \\
LR(1)   & 76.78 & 95.16 & 2.15 & 0.464 & 0.349 & 0.939 & 1.13 & 0.712 & 0.352 \\
\end{tabular}
\label{t:cc}
\end{ruledtabular}
\end{table}
%


\newpage

\newpage

\end{document}